\documentclass[a4paper,11pt]{article}
\usepackage{pos}
\usepackage{physics}
\usepackage{slashed}
\usepackage{xcolor,graphicx}
\usepackage{subfigure}
\usepackage{etoolbox}
\usepackage{bbold}
\usepackage{xspace}
\usepackage{appendix}

\usepackage{textpos}
\usepackage{slashed}
\usepackage{stackrel}
\usepackage{graphicx}
\usepackage{adjustbox}
\usepackage{tcolorbox}
\usepackage[framemethod=tikz]{mdframed}

\newcommand{\intx}{\int d^4x}
\newcommand{\Dintx}{\int d^Dx}

\allowdisplaybreaks

\title{Advances at the $\gamma_5$-Frontier}

\author[a]{Paul Kühler}
\author[a]{Dominik  Stöckinger}
\author[a]{Matthias Weißwange}

\affiliation[a]{Institut für Kern- und Teilchenphysik,\\
  Zellescher Weg 19, DE-01069 Dresden, Germany}

\emailAdd{paul.kuehler@tu-dresden.de}
\emailAdd{dominik.stoeckinger@tu-dresden.de}
\emailAdd{matthias.weisswange@tu-dresden.de}

\abstract{These proceedings discuss the current progress
of the no-compromise approach to the dimensional renormalization 
of chiral gauge theories in the context of the BMHV scheme with non-anticommuting $\gamma_5$.
Despite spuriously breaking BRST-invariance in intermediate steps, it
is the only scheme which handles the well-known $\gamma_5$-problem
mathematically consistently.
We begin with a brief motivation, followed by an exposition of our methodology.
Specifically, we illustrate the symmetry restoration procedure to obtain
the required symmetry-restoring counterterms 
and provide insights into
our computational setups, including recent developments.
Building up on this, we present recent results and advances for the 
multi-loop renormalization of Abelian and non-Abelian chiral gauge theories, before concluding 
this article with a discussion on prospective 
implementations of the BMHV regularization in SM-like models.}

\FullConference{Loops and Legs in Quantum Field Theory (LL2024)\\
 14-19, April, 2024\\
Wittenberg, Germany\\}

\tableofcontents

\begin{document}
\maketitle
\newpage
\section{Introduction}
Dimensional regularization of chiral gauge theories is affected by the well-known $\gamma_5$ problem. 
The three properties
\begin{itemize}
    \item[$(i)$] anticommutativity with $\gamma^\mu$, 
    \item[$(ii)$] non-zero $\text{Tr}(\gamma_5\gamma^\mu\gamma^\nu\gamma^\rho\gamma^\sigma)$,
    \item[$(iii)$] cyclicity of traces,
\end{itemize}
become inconsistent in $D\ne4$ dimensions,
e.g.\ $(i)$ and $(iii)$ imply that the trace in $(ii)$ is zero.

There is a multitude of proposals how to
define a $D$-dimensional continuation of $\gamma_5$ and many of them
are routinely applied in practical computations (for a review see
e.g.\ Ref.\ \cite{Jegerlehner:2000dz}).

A mathematically rigorous scheme is provided by the original proposal of
Ref.\ \cite{tHooft:1972tcz}, which was later further formalized in
Ref.\ \cite{Breitenlohner:1975hg} --- the BMHV scheme. Its key advantage is that full mathematical
consistency and complete all-order proofs are established
\cite{Breitenlohner:1975hg}, see also the review \cite{Belusca-Maito:2023wah}. Here we focus on the BMHV scheme, accept its practical difficulties, and aim to work out how to use it in chiral gauge theories of phenomenological importance.
Specifically, we aim to provide the
required symmetry-restoring counterterms which compensate the spurious
breaking of gauge invariance caused by the non-anticommuting
$\gamma_5$. For an introduction to the key ideas and methods as well as past results we refer to our articles \cite{Belusca-Maito:2020ala,Belusca-Maito:2021lnk,Belusca-Maito:2022wem,Stockinger:2023ndm}, the proceedings of Loops\&Legs 2022 \cite{Belusca-Maito:2022usw} and our review \cite{Belusca-Maito:2023wah}.
In the present proceedings we provide a brief summary of the relevant definitions and methods (Secs.\  \ref{Sec:Gamma5Problem}, \ref{Sec:SymmetryRestorationProcedure} and \ref{Sec:ComputationalSetup}), then we report on progress in three directions: progress in the number of loops (Sec.\ \ref{Sec:AbelianMultiLoop}), progress in treating non-Abelian chiral gauge theories (Sec.\ \ref{Sec:NonAbelianTheories}), and progress towards specific details of the electroweak Standard Model (Sec.\ \ref{Sec:ExplorationsAndSM}).

\section{Non-anticommuting $\gamma_5$ in the BMHV Scheme}\label{Sec:Gamma5Problem}

In the BMHV scheme, the formally $D=(4-2\epsilon)$-dimensional space is viewed as a direct sum of
$4$-dimensional and $(D-4)$-dimensional subspaces, with associated metric tensors 
$g^{\mu\nu}=\overline{g}^{\mu\nu}+\widehat{g}^{\mu\nu}$. These metric tensors act as projection operators onto the respective 
spaces such that e.g.\ $\overline{g}^{\mu\nu}\widehat{g}_{\nu\rho}=0$. 
Using these tensors, any formally
$D$-dimensional quantity $k^\mu$ can be similarly split as
\begin{align}\label{Eq:SplitOfDdimk}
  k^\mu &= \overline{k}^\mu + \widehat{k}^\mu \,.
\end{align}
The split can be done for objects such as momentum vectors,
gauge fields, metric tensors, and in particular for $\gamma^\mu$
matrices, $\gamma^\mu = \overline{\gamma}^\mu+\widehat{\gamma}^\mu$.

In the BMHV scheme, the matrix $\gamma_5$  is defined as an
intrinsically 4-dimensional object. It satisfies
  \begin{align}
\label{eq:Gamma5DReg_A}
    \{\gamma_5, \overline{\gamma}^\mu\} &= 0 \, , &
    [\gamma_5, \widehat{\gamma}^\mu] &= 0 \, ,
  \end{align}
and thus it breaks full $D$-dimensional Lorentz covariance. The
usual anticommutation relation holds only for the purely $4$-dimensional
parts of the $\gamma^\mu$ matrices. Importantly, this definition is
consistent with 
the cyclicity of traces and with the relation $\gamma_5 = \frac{-i}{4!}
\epsilon_{\mu\nu\rho\sigma} \overline{\gamma}^\mu \overline{\gamma}^\nu \overline{\gamma}^\rho
\overline{\gamma}^\sigma$. 

Due to these modified algebraic relations, the BMHV scheme leads to a
regularization induced, spurious symmetry breaking.
To explain the problem we consider a simple $U(1)$ chiral gauge theory 
with a set of fermion fields $\psi_i$. We consider only the right-handed fermions to interact with the gauge field $A^\mu$ via  ``hypercharges''
$\mathcal{Y}_R{}_i$. The corresponding fermionic part of the $D$-dimensional Lagrangian can
be written as
\begin{align}
	\mathcal{L}_\text{fermions} &= i \overline{\psi}_i \slashed{\partial} {\psi}_i - e \mathcal{Y}_R{}_{i} \overline{\psi_R}_i \slashed{A} {\psi_R}_i \, .
\label{Eq:LfermionsR}  
\end{align}
Here $\psi_R=P_R\psi$ with the right-chiral projector
$P_R=(1+\gamma_5)/2$. Importantly, the kinetic term must involve the
full, $D$-dimensional derivative $\slashed{\partial}$ in order to
generate a regularized, $D$-dimensional propagator denominator in
Feynman diagrams. 
The mismatch between the $D$-dimensional kinetic term $\overline{\psi}\slashed{\partial}{\psi}$
and the $4$-dimensional interaction current $\overline{\psi_R}\slashed{A}{\psi_R}=\overline{\psi_R}\overline{\slashed{A}}{\psi_R}$ 
causes a breaking of gauge invariance in
$D$-dimensions.

Technically, in the quantized theory, gauge invariance is replaced by BRST invariance
involving the Faddeev-Popov ghost field $c$. BRST invariance manifests itself in terms of
Ward and Slavnov-Taylor identities which must be satisfied by renormalized Green functions. These can be summarized by the
expression 
\begin{equation}\label{Eq:SlavnovTaylorIdentity}
\mathcal{S}(\Gamma)=\intx\,\frac{\delta\Gamma}{\delta\phi_i(x)}\frac{\delta\Gamma}{\delta K_{\phi_i}(x)}=0,
\end{equation}
where $\mathcal{S}$ is the
Slavnov-Taylor operator and $\Gamma$ the renormalized, finite
generating functional of 1PI Green functions. In this formalism, the breaking of gauge
invariance of the regularized Lagrangian causes a
non-zero result of the $D$-dimensional Slavnov-Taylor operator applied to the
classical action $S_0$ in $D$ dimensions,
\begin{align}\label{DeltahatDef}  
    \mathcal{S}_D(S_0) &= 
    \widehat{\Delta} \, \equiv 
    - \Dintx \,
    (e \mathcal{Y}_R{}_{i}) c 
    \left\{
    \overline{\psi}_i \left(\overset{\leftarrow}{\widehat{\slashed{\partial}}} \mathbb{P}_{\mathrm{R}} 
    + 
    \overset{\rightarrow}{\widehat{\slashed{\partial}}} \mathbb{P}_{\mathrm{L}} \right) \psi_i
    \right\}
    \, .
\end{align}
This non-zero result defines a composite operator $\widehat{\Delta}$, which is evanescent, i.e.\ which vanishes in purely 4 dimensions. At the loop level, this evanescent tree-level breaking of the Slavnov-Taylor identity results in non-evanescent breakings of the Slavnov-Taylor identity. Such breakings must be cancelled by suitable symmetry-restoring counterterms.


\section{Symmetry Restoration Procedure}\label{Sec:SymmetryRestorationProcedure}

Following from the previous section, the renormalization of chiral gauge theories
within the BMHV scheme includes a symmetry restoration procedure, determining the
aforementioned symmetry-restoring counterterms, in order to cancel spurious
symmetry breakings induced by the regularization, such that the Slavnov-Taylor
identity is satisfied after renormalization, i.e.\
\begin{equation}\label{Eq:UltimateSTI}
    \begin{aligned}
        \mathop{\text{LIM}}_{D \, \to \, 4} \, (\mathcal{S}_D(\Gamma_\mathrm{DRen})) = 0,
    \end{aligned}
\end{equation}
where $\Gamma_\mathrm{DRen}$ is the fully renormalized, $D$-dimensional 1PI effective action and where the $\mathop{\text{LIM}}_{D \, \to \, 4}$ operation includes not only setting $D=4$ but also dropping evanescent objects such as $\widehat{g}^{\mu\nu}$.

In principle, there are two different ways to obtain the symmetry-restoring counterterms.
One way would be to determine all Green functions in $\mathcal{S}_{D}(\Gamma^{(n)}_\text{subren}+S^n_\text{sct})$,
including their finite parts, and determine potentially non-vanishing breakings, i.e.\ test
the validity of the Slavnov-Taylor identity.
This is an obvious and straightforward strategy, which operates on ordinary Green functions.
However, it is unnecessarily complicated and lacks in efficiency, because there are in principle 
infinitely many identities between Green functions and it involves the computation 
of finite and non-local contributions. In particular, non-local finite contributions
cannot contribute to the symmetry violation and will thus drop out of the Slavnov-Taylor 
identities.
The calculation of such non-local contributions is a major bottleneck at the multi-loop
level.

Another approach makes use of the regularized quantum action principle of DReg (see Ref.\ \cite{Breitenlohner:1977hr} and 
also the review \cite{Belusca-Maito:2023wah}),
\begin{equation}\label{Eq:QAPofDReg}
    \begin{aligned}
        \mathcal{S}_D(\Gamma_\mathrm{DRen})=(\widehat{\Delta}+\Delta_{\mathrm{ct}})\cdot\Gamma_\mathrm{DRen},
    \end{aligned}
\end{equation}
where a possible symmetry breaking is rewritten as an operator insertion into the
1PI effective action, with composite operator
\begin{equation}\label{Eq:DefinitionOfDelta}
    \begin{aligned}
        \Delta \equiv \widehat{\Delta} + \Delta_{\mathrm{ct}} = \mathcal{S}_D(S_{0} + S_{\mathrm{ct}}),
    \end{aligned}
\end{equation}
representing the symmetry violation and whose lowest-order part is evanescent,
a property which turns out to be essential for higher-order calculations, being the root of the efficiency of this method.
Plugging Eq.\ (\ref{Eq:QAPofDReg}) into Eq.\ (\ref{Eq:UltimateSTI}) leads to the perturbative
requirement
\begin{equation}\label{Eq:PerturbativeSTIRequirement}
    \begin{aligned}
        \mathop{\text{LIM}}_{D \, \to \, 4} \, \bigg(\widehat{\Delta}\cdot\Gamma_\mathrm{DRen}^n+\sum_{k=1}^{n-1}\Delta^k_\mathrm{ct}\cdot\Gamma^{n-k}_\mathrm{DRen}+\Delta^n_\mathrm{ct}\bigg)=0,
    \end{aligned}
\end{equation}
with $n$ indicating the loop order, which is then used as
the starting point of the iterative symmetry restoration procedure.
Hence, according to Eq.\ (\ref{Eq:PerturbativeSTIRequirement}), 
this approach then requires the calculation of $\Delta$-operator inserted
Green functions order-by-order in perturbation theory
to extract the symmetry violation.
This leads to a significant simplification,
since only the UV-divergent part of power-counting divergent, 1PI Green functions
of this kind needs to be computed, which is due to the evanescence property 
of $\Delta$.
In a nutshell, the finite, symmetry-violating contributions, necessary for the
symmetry-restoring counterterms, are local and emerge from UV-divergences of
such $\Delta$-operator inserted Green functions, as the evanescent part
of $\Delta$ hits a $1/\epsilon$-pole.
This feature turns out be crucial at higher loop orders and highlights the main
advantage of this method, i.e.\ its much higher efficiency compared to the method 
presented first.
A further advantage is that, in general, there are fewer diagrams with $\Delta$-insertion
than ordinary ones.

Due to the significantly higher efficiency,
we primarily use the latter method employing the regularized quantum action principle of DReg. The former, more straightforward method is used for cross-checks 
at the 1-loop and 2-loop level.
For a more detailed description of both methods and a comprehensive example comparing 
both methods using the gauge boson self-energy in an Abelian chiral gauge theory
with fermionic content given by Eq.\ (\ref{Eq:LfermionsR}), the reader is referred 
to the review article \cite{Belusca-Maito:2023wah}.


\section{Computational Setup and calculational Methods}\label{Sec:ComputationalSetup}

Currently, we have three independent computational setups in charge to perform loop calculations. 
The first two are \texttt{Mathematica}-based setups, where
most computations are performed in \texttt{Mathematica} \cite{mathematica}.
In particular, Feynman diagrams are generated using 
\texttt{FeynArts} \cite{Hahn:2000kx} and most symbolic manipulations,
especially those related to the Dirac algebra, are performed with the help of
\texttt{FeynCalc} \cite{Mertig:1990an, Shtabovenko:2016sxi, Shtabovenko:2020gxv, Shtabovenko:2021hjx}.
The integral reduction is performed with the \texttt{C++} version of the
software \texttt{FIRE} \cite{Smirnov:2019qkx}, which
is interfaced to \texttt{Mathematica} using 
the package \texttt{FeynHelpers} \cite{Shtabovenko:2016whf}.
These two setups have successfully been tested and used at the 3-loop level for
Abelian chiral gauge theories, see Ref.\ \cite{Stockinger:2023ndm}, and 
at the 2-loop level for non-Abelian chiral gauge theories (work in progress).

However, already in these use cases, performance limitations within the 
\texttt{Mathematica}-based setups are evident, with \texttt{Mathematica} notably acting as the bottleneck, 
especially in the context of evaluating the Dirac algebra with many $\gamma$-matrices, 
performing the tensor reduction with high tensor ranks and in general processing 
a large amount of terms.
In order to address these challenges and expand both the number of loops 
(e.g.\ up to 4-loop applications) and the complexity of the theories considered 
(e.g.\ the Standard Model), we are currently developing a third setup based on 
\texttt{FORM} \cite{Vermaseren:1991,Vermaseren:2000nd,Kuipers:2012rf,Ruijl:2017dtg}.
In this framework, all Feynman diagrams are generated using \texttt{QGRAF} \cite{Nogueira:1991ex}.
Further, we employ \texttt{Feynson} \cite{Maheria:2022dsq} and \texttt{Reduze2} \cite{vonManteuffel:2012np} 
to identify integral family and sector symmetries,
whereas the actual IBP-reduction is carried out using the \texttt{C++} version of 
\texttt{FIRE}. 
In particular, we use \texttt{Feynson} to map all integrals to a minimal set
of integral families and \texttt{Reduze2} to produce symmetry relations between
integrals, which are then fed into \texttt{FIRE}, enabling it to reduce all 
integrals to a minimal set of preferred master integrals.
\texttt{Mathematica} is still used for interfacing between various software tools, 
and for the automated generation of \texttt{FORM} code.
This setup significantly enhances the computational performance, 
making it very promising for future Standard Model applications.

In any setup, the computational effort required for the renormalization of chiral 
gauge theories within the BMHV scheme is greater than that for vector-like gauge 
theories due to the following two challenges.

First, due to the regularization induced symmetry breaking, Ward or Slavnov-Taylor 
identities cannot be used to circumvent the calculation of multi-leg 1PI Green functions
as it is usually done 
(see e.g.\ \cite{Davies:2019onf,Davies:2021mnc,Herren:2021vdk,Herzog:2017ohr,Luthe:2017ttg}).
Thus, 1PI Green functions up to and including five external legs need to be calculated
to fully renormalize a chiral gauge theory at a given loop order.

Second, because of the regularization induced symmetry breaking due to the BMHV algebra, see Eqs.\ (\ref{Eq:SplitOfDdimk}) and (\ref{eq:Gamma5DReg_A}),
there are also more Lorentz covariants 
than usual, cf.\ results in Refs.\ \cite{Belusca-Maito:2023wah,Belusca-Maito:2020ala,Belusca-Maito:2021lnk,Stockinger:2023ndm}, 
making it more difficult to construct simple projectors.
Moreover, one cannot simply contract the integrand with non-$D$-dimensional projectors 
or perform the Dirac algebra, in order to
lower the tensor rank, \textit{before} expressing all numerator polynomials of
loop momenta in terms of a linear combination of inverse propagator denominators, and thus rewriting the integrand
in the well-known ``index-form'', i.e.\ $I(n_1,n_2,\ldots,n_z)$, used as input for the IBP-reduction.
This is due to the fact that such contractions can lead to $4$- and $(D-4)$-dimensional 
loop-momentum combinations in the numerator, which cannot be expressed via the $D$-dimensional 
propagator denominators.
Therefore, this procedure and consequently the tensor reduction must be performed prior to any 
contractions with non-$D$-dimensional quantities, including the evaluation of the Dirac algebra. 
Specifically, the tensor reduction thus requires the application of generic $D$-dimensional projectors 
acting on tensors of relatively high ranks.
Note that this obstacle can be circumvented if one can treat such numerator-polynomials
containing $4$- and $(D-4)$-dimensional loop-momenta.
A method to achieve this, in fact, has been introduced in Ref.\ \cite{Bern:1995db}, 
at the 1-loop level, 
via so-called ``$\mu$-term'' insertions of the form $\widehat{k}_i \cdot \widehat{k}_j$.
Such $\mu$-term inserted Feynman integrals can then be rewritten in terms of standard
dimensionally-shifted ones admitting the same propagator structure.
This algorithm has been extended to the 2-loop level in Ref.\ \cite{Bern:2002tk} 
and reviewed, as well as applied, in Ref.\ \cite{Heller:2020owb}.
While we and the authors of Ref.\ \cite{Heller:2020owb} believe that this method 
can be extended to higher loop orders, which are necessary for our purposes, 
additional effort is required. 
Currently, we have not implemented this method and postpone any potential discussion 
to future research efforts.

To address both challenges, we employ a tadpole decomposition, explained below,
in order to map all Feynman integrals to fully massive single-scale vacuum bubble integrals,
such that possible IR-divergences are avoided and the computational complexity is reduced 
drastically. 
The simplification is evident not only in the IBP-reduction of the integrals,
due to the presence of only one remaining integral family with one 
remaining scale in the problem, and the existence 
of solutions for the master integrals, enabling calculations at loop orders $\geq3$,
but also in the tensor reduction procedure, as possible tensor structures
are completely restricted to metric tensors $g_{\mu\nu}$.
However, for tensor integrals of rank $8$ and higher, tensor reduction 
remains a significant computational challenge. 
Although tensors of rank $8$ are manageable with \texttt{FeynCalc},
the performance is suboptimal. 
Higher-rank tensors 
require more efficient implementations that utilize symmetries.
We addressed this issue in our third setup based on \texttt{FORM}, primarily based 
on the methods described in Refs.\
\cite{Herzog:2017ohr,Ruijl:2018poj}, where we precomputed 
tensor reduction tables up to rank 16.

Now, we briefly discuss some details of the 
tadpole decomposition method, 
which plays a crucial role in our 
computations.
Counterterms are local polynomials in external momenta and 
(for mass-independent schemes) internal masses.
Hence, we can extract the UV-divergences utilizing an 
infrared rearrangement to achieve this task.
Specifically, we use a tadpole decomposition, first introduced in Refs.\
\cite{Misiak:1994zw,Chetyrkin:1997fm}, where the infrared rearrangement is 
realized via the
following exact decomposition
\begin{equation} \label{Eq:StandardTadpoleExpansion}
    \begin{aligned}
        \frac{1}{(k+p)^2} = \frac{1}{k^2 - M^2} - \frac{p^2 + 2 \, k \cdot p + M^2}{k^2 - M^2} \, \frac{1}{(k+p)^2},
    \end{aligned}
\end{equation}
where $k$ is a loop momentum or any linear combination of loop momenta.
After recursively applying this decomposition sufficiently often
and subsequently truncating the last term, a propagator with 
external momentum is decomposed into one or more without external momentum, 
but with an auxiliary mass scale $M$.

Since this form of the tadpole decomposition method is not convenient 
for computer implementations because of
subtleties in higher-order applications w.r.t.\ the arrangement of
the momentum-routing of genuine $L$-loop and associated $(<\!L)$-loop 
counterterm-inserted
Feynman diagrams due to subdivergences, 
we opted for an improved tadpole expansion
(see Ref.\ \cite{Stockinger:2023ndm}), as implied in 
Refs.\ \cite{Misiak:1994zw,Chetyrkin:1997fm} and described in more
detail in Ref.\ \cite{Lang:2020nnl}.
Here, the auxiliary mass scale $M$ is introduced in every propagator,
followed by a Taylor-expansion in external momenta 
(and in internal/physical masses if present). 
For example, for a massless propagator, we obtain
\begin{equation}
    \begin{aligned}
        \frac{1}{(k+p)^2} \longrightarrow \frac{1}{(k+p)^2 - M^2} = \frac{1}{k^2 - M^2} - \frac{p^2 + 2 \, k \cdot p}{(k^2 - M^2)^2}  + \frac{(p^2 + 2 \, k \cdot p)^2}{(k^2 - M^2)^3} + \ldots,
    \end{aligned}
\end{equation}
where it can be seen that the same result as with the exact decomposition is obtained 
when neglecting numerator terms $\propto M^2$ in Eq.\ (\ref{Eq:StandardTadpoleExpansion}).
However, omitting these numerator terms $\propto M^2$ requires a compensation, 
particularly at the multi-loop level where subdivergences arise.
This is accomplished by constructing and incorporating all 
possible auxiliary counterterms proportional to $M^2$
at a given order. 
Both the auxiliary mass $M^2$ and the auxiliary counterterms
$\propto M^2$ are used 
solely for the evaluation of the Feynman integrals
and are not intrinsic to the theory;
thus, they can be regarded as a mathematical tool.
Specifically, the auxiliary gauge boson mass counterterm does not cause 
any issues regarding gauge invariance.

To conclude this section, we address some intricacies
regarding the implementation of counterterms in our different setups.
In our \texttt{Mathematica}-based setups, challenges
arise primarily in \texttt{FeynArts} for counterterms containing external sources
(see Sec.\ \ref{Sec:NonAbelianTheories}), while 
difficulties in our \texttt{FORM}-based setup mainly stem
from the implementation of counterterms related to propagators.
Concretely, this is realised in \texttt{FORM} by recursively applying the 
following modified propagators
\begin{align}
    \begin{split}\label{Eq:RecursiveFermionPropagator}
        \mathcal{D}_{f,ij}(k) &= \frac{i \slashed{k}}{k^2} \Big( \delta_{ij} 
        + i \big[
        \delta \overline{Z}_{f,\mathrm{R},ik} \, \mathbb{P}_{\mathrm{L}} \slashed{k} \mathbb{P}_{\mathrm{R}} 
        + \delta \overline{Z}_{f,\mathrm{L},ik} \, \mathbb{P}_{\mathrm{R}} \slashed{k} \mathbb{P}_{\mathrm{L}}\\
        &\qquad\qquad\,\, + \delta \widehat{Z}_{f,\mathrm{LR},ik} \, \mathbb{P}_{\mathrm{R}} \slashed{k} \mathbb{P}_{\mathrm{R}}
        + \delta \widehat{Z}_{f,\mathrm{RL},ik} \, \mathbb{P}_{\mathrm{L}} \slashed{k} \mathbb{P}_{\mathrm{L}}
        \big] \mathcal{D}_{f,kj}(k) \Big),
    \end{split}\\[1.5ex]
    \begin{split}\label{Eq:RecursiveGaugeBosonPropagator}
        \mathcal{D}_g^{\mu\nu}(k) &= \frac{-i}{k^2} \Big(g^{\mu\rho}- \big(1 - \xi\big) \frac{k^{\mu}k^{\rho}}{k^2}\Big) 
        \Big( \delta_{\rho}^{\phantom{\rho}\nu} - i \Big[ 
        \delta\overline{Z}_{M} M^2 \overline{g}_{\rho\sigma} 
        + \delta\widehat{Z}_{M} M^2 \widehat{g}_{\rho\sigma}\\
        &\hspace{1.81cm} + \delta Z^{D}_g \big( g_{\rho\sigma}k^2 - k_{\rho}k_{\sigma} \big) 
        + \delta \overline{Z}^4_g \big( \overline{g}_{\rho\sigma}\overline{k}^2 - \overline{k}_{\rho}\overline{k}_{\sigma} \big)\\
        &\hspace{1.81cm} + \delta Z_g^{aa} \, \overline{g}_{\rho\sigma}\overline{k}^2
        + \delta Z_g^{ab} \, \overline{g}_{\rho\sigma}\widehat{k}^2
        + \delta Z_g^{ba} \, \widehat{g}_{\rho\sigma}\overline{k}^2
        + \delta Z_g^{bb} \, \widehat{g}_{\rho\sigma}\widehat{k}^2\\
        &\hspace{1.81cm} - \delta Z_g^{cc} \, \overline{k}_{\rho}\overline{k}_{\sigma}
        - \delta Z_g^{cd} \, \overline{k}_{\rho}\widehat{k}_{\sigma}
        - \delta Z_g^{dc} \, \widehat{k}_{\rho}\overline{k}_{\sigma}
        - \delta Z_g^{dd} \, \widehat{k}_{\rho}\widehat{k}_{\sigma}
        \Big] \mathcal{D}_g^{\sigma\nu}(k) \Big),
    \end{split}
\end{align}
up to the necessary power in $\hbar$.
This is analogous to the approach taken by the authors of Ref.\ \cite{Luthe:2017ttg},
but adapted to serve our purposes within the framework of the BMHV scheme, 
where the counterterm contributions in the propagators contain all possible Lorentz covariants.
Finally, it is important to note that the $\delta Z$'s do not generally result from a multiplicative 
renormalization transformation,
as they also cover symmetry-restoring counterterms.


\section{Abelian chiral Gauge Theory at the 3-loop Level}\label{Sec:AbelianMultiLoop}

Using the previously discussed methodology, we have fully
renormalized an Abelian chiral gauge theory with purely right-handed interaction defined by the fermionic Lagrangian in Eq.\ (\ref{Eq:LfermionsR}) at the 3-loop level. These results currently represent the highest-loop application of the BMHV scheme to a chiral gauge theory, and they can be found in Ref.\ \cite{Stockinger:2023ndm}, along with an extensive discussion. Here we present a brief review 
and outline future research efforts. 
These results have been obtained using a \texttt{Mathematica}-based computational setup,
as described in Sec.\ \ref{Sec:ComputationalSetup}, and the tadpole decompostion
illustrated therein.
To present the results concisely, we define coefficients 
whose explicit forms are provided in appendix \ref{App:3-Loop-Coeffs}.

As an example of an explicit 3-loop result we present,
\begin{equation}\label{Eq:GaugeBosonSelfEnergy3Loop}
    \begin{aligned}
        i \widetilde{\Gamma}_{AA}^{\nu\mu}(p) \big|_{\text{div}}^{3} = 
        &- \frac{ie^6}{(16 \pi^2)^3} \, 
        \bigg[ \mathcal{B}_{AA}^{3, \text{inv}} \, \frac{1}{\epsilon^2} + \mathcal{A}_{AA}^{3, \text{inv}} \, \frac{1}{\epsilon} \bigg]
        \Big(\overline{p}^{\mu} \overline{p}^{\nu} - \overline{p}^2 \overline{g}^{\mu\nu}\Big)\\
        &- \frac{ie^6}{(16 \pi^2)^3} \,
        \bigg[ \widehat{\mathcal{C}}_{AA}^{3, \text{break}} \, \frac{1}{\epsilon^3} + \widehat{\mathcal{B}}_{AA}^{3, \text{break}} \, \frac{1}{\epsilon^2} + \widehat{\mathcal{A}}_{AA}^{3, \text{break}} \, \frac{1}{\epsilon} \bigg] \, \widehat{p}^2 \, \overline{g}^{\mu\nu}\\
        &+ \frac{ie^6}{(16 \pi^2)^3} \, \overline{\mathcal{A}}_{AA}^{3, \text{break}} \, \frac{1}{\epsilon} \, \overline{p}^2 \, \overline{g}^{\mu\nu},
    \end{aligned}
\end{equation}
which is the UV-divergent contribution to the gauge boson self-energy with the familiar transversal contribution shown in the first line and 
additional symmetry-breaking contributions in the last two lines.
The symmetry violation arising from the regularization scheme is local
and can thus be cancelled by symmetry-restoring counterterms. In order to determine the symmetry breaking at the finite level we could in principle fully compute the gauge boson self-energy (corresponding to the first method discussed in Sec.\ \ref{Sec:SymmetryRestorationProcedure}). This would however require different calculational techniques. In order to extract finite symmetry-breaking contributions within
the framework of the tadpole decomposition, we employ the regularized
quantum action principle of DReg (the second method explained in Sec.\ 
\ref{Sec:SymmetryRestorationProcedure}).
We start from Eq.\ (\ref{Eq:PerturbativeSTIRequirement})
with $n=3$, indicating the loop-order at hand, and compute 
the contributing operator-inserted Green functions.

First, the lowest-order
component of $\Delta$, the evanescent 
$\widehat{\Delta}$-operator representing the tree-level breaking,
needs to be inserted into the subrenormalized 3-loop
1PI effective action, giving rise to $\widehat{\Delta}$-inserted diagrams such as
\begin{equation}
    \begin{tabular}{rl}
                $\begin{aligned}
                    i \big( \widehat{\Delta} \cdot \widetilde{\Gamma}_{\mathrm{DRen}}^3 \big)_{A_{\mu} c} =
                \end{aligned}$
                &\raisebox{-40pt}{\includegraphics[trim={0.15cm 0.02cm 0.03cm 0.015cm},clip,scale=0.3]{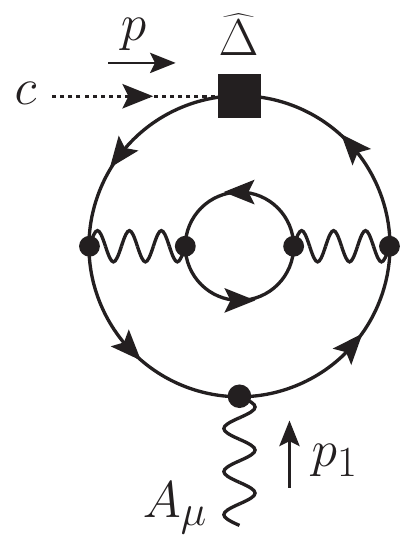}}
                $\begin{aligned}
                    +
                \end{aligned}$
                \raisebox{-40pt}{\includegraphics[trim={0.03cm 0.02cm 0.03cm 0.015cm},clip,scale=0.3]{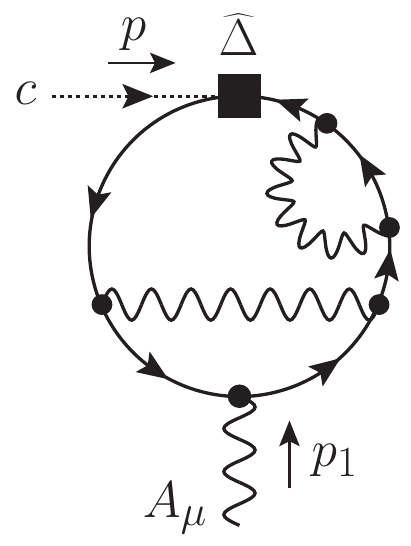}}
                $\begin{aligned}
                    +
                \end{aligned}$
                \raisebox{-40pt}{\includegraphics[trim={0.03cm 0.02cm 0.03cm 0.015cm},clip,scale=0.3]{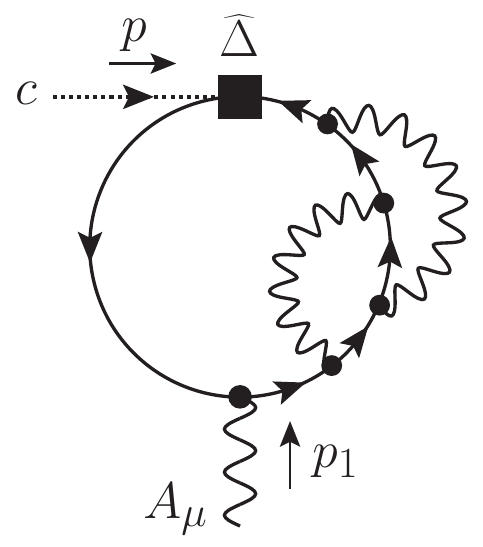}}
                $\begin{aligned}
                    + \ldots
                \end{aligned}$
                $\begin{aligned}
                    +
                \end{aligned}$
                \raisebox{-40pt}{\includegraphics[trim={0.03cm 0.02cm 0.03cm 0.015cm},clip,scale=0.3]{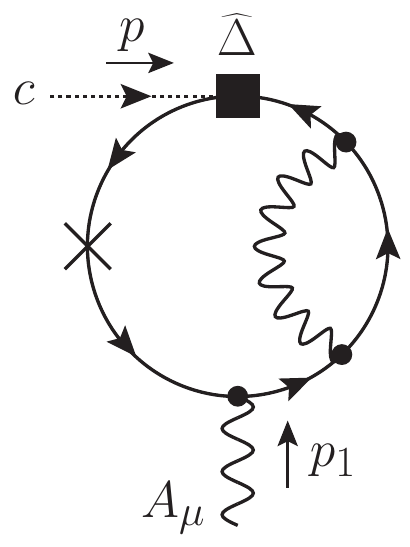}}\\
                $\begin{aligned}
                    +
                \end{aligned}$
                &\raisebox{-40pt}{\includegraphics[trim={0.15cm 0.02cm 0.03cm 0.015cm},clip,scale=0.3]{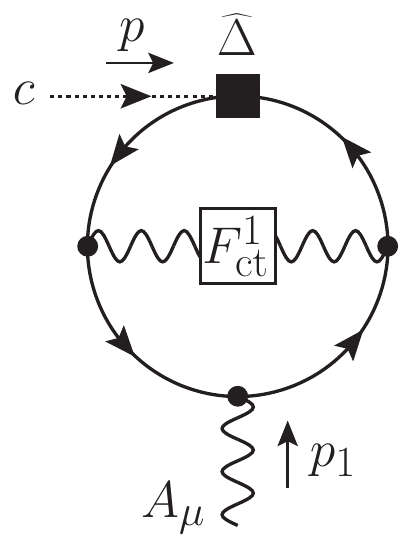}}
                $\begin{aligned}
                    +
                \end{aligned}$
                \raisebox{-40pt}{\includegraphics[trim={0.03cm 0.02cm 0.03cm 0.015cm},clip,scale=0.3]{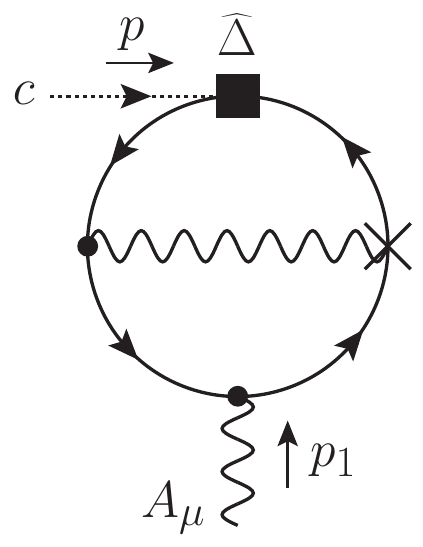}}
                 $\begin{aligned}
                    + \ldots
                \end{aligned}$
                $\begin{aligned}
                    +
                \end{aligned}$
                \raisebox{-40pt}{\includegraphics[trim={0.03cm 0.02cm 0.03cm 0.015cm},clip,scale=0.3]{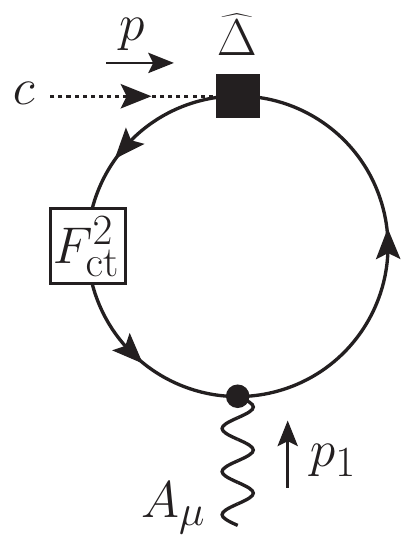}}
                $\begin{aligned}
                    +
                \end{aligned}$
                \raisebox{-40pt}{\includegraphics[trim={0.03cm 0.02cm 0.03cm 0.015cm},clip,scale=0.3]{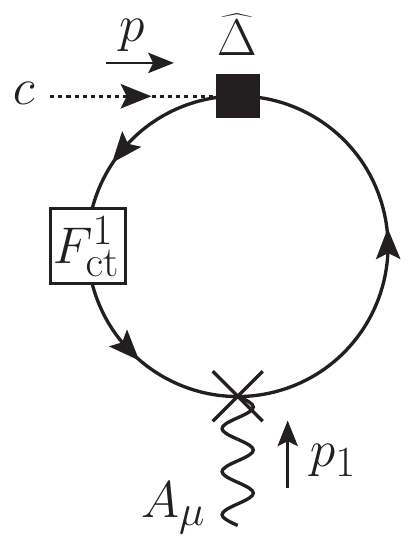}}\\
                 $\begin{aligned}
                    +
                \end{aligned}$
                &$\ldots$\phantom{\raisebox{-31pt}{\includegraphics[trim={0.01cm 0.02cm 0.01cm 0.01cm},clip,scale=0.3]{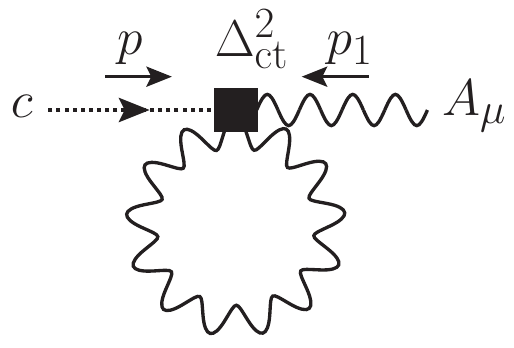}}}
    \end{tabular}\nonumber
    \vspace{-0.5cm}
\end{equation}
which contribute to the breaking of the gauge boson self-energy.
As usual one needs to calculate not only the genuine 3-loop diagrams,
but also diagrams with counterterm insertions of lower loop-order,
as illustrated above.
It is important to note that in addition to 
the standard counterterms marked with $\times$, there are also
finite-symmetry restoring counterterms of lower loop-orders 
indicated by $F_{\mathrm{ct}}^{n-k}$.

Second, the higher-order components $\Delta_{\mathrm{ct}}^{k}$
of $\Delta$ need to be inserted into the subrenormalized
1PI effective action of order $n-k$, giving rise to
\begin{equation}
    \begin{tabular}{rl}
                $\begin{aligned}
                    i \big( \Delta^1_{\text{ct}} \cdot \widetilde{\Gamma}_{\mathrm{DRen}}^2 \big)_{A_{\mu} c} =
                \end{aligned}$
                &\raisebox{-40pt}{\includegraphics[trim={0.03cm 0.02cm 0.03cm 0.015cm},clip,scale=0.3]{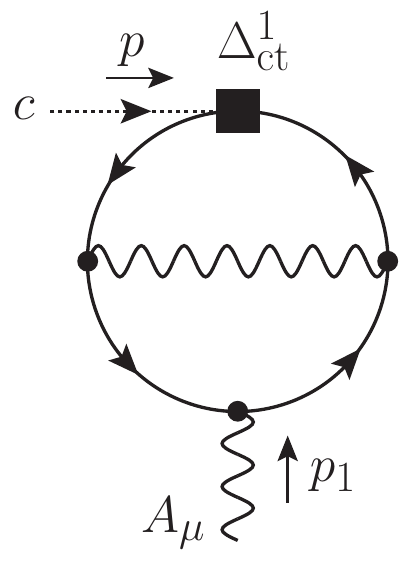}}
                $\begin{aligned}
                    +
                \end{aligned}$
                \raisebox{-40pt}{\includegraphics[trim={0.03cm 0.02cm 0.03cm 0.015cm},clip,scale=0.3]{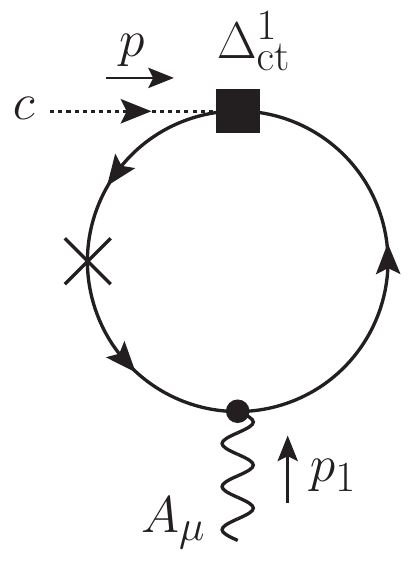}}
                $\begin{aligned}
                    + \ldots
                \end{aligned}$
    \end{tabular}\nonumber
\end{equation}
\begin{equation}
    \begin{tabular}{rl}
                $\begin{aligned}
                i \big( \Delta^2_{\text{ct}} \cdot \widetilde{\Gamma}_{\mathrm{DRen}}^1 \big)_{A_{\mu} c} =
                \end{aligned}$
                &\raisebox{-40pt}{\includegraphics[trim={0.03cm 0.02cm 0.03cm 0.015cm},clip,scale=0.3]{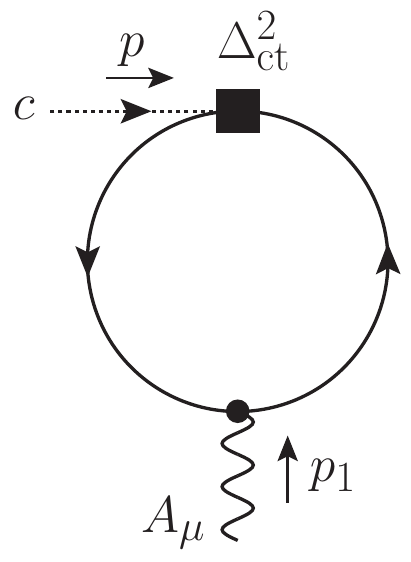}}
                $\begin{aligned}
                    +
                \end{aligned}$
                \raisebox{-31pt}{\includegraphics[trim={0.01cm 0.02cm 0.01cm 0.01cm},clip,scale=0.3]{images_MW/cA1Lct2_8.pdf}}.
    \end{tabular}\nonumber
\end{equation}
These higher-order contributions $\Delta_{\mathrm{ct}}^{k}$ 
represent the breaking of the order $k$ and are obtained
via Eq.\ (\ref{Eq:DefinitionOfDelta}) order-by-order 
in perturbation theory.

Summing up all three contributions, we obtain
\begin{equation}\label{Eq:GaugeBosonSelfEnergy3LoopBreaking}
    \begin{aligned}
        i \big( \Delta \cdot \widetilde{\Gamma} \big)_{A_{\mu} c}^{3} = &\phantom{-} i \big( \widehat{\Delta} \cdot \widetilde{\Gamma}^3 \big)_{A_{\mu} c} + i \big( \Delta^1_{\text{ct}} \cdot \widetilde{\Gamma}^2 \big)_{A_{\mu} c} + i \big( \Delta^2_{\text{ct}} \cdot \widetilde{\Gamma}^1 \big)_{A_{\mu} c}\\
        = &- \frac{e^6}{(16 \pi^2)^3} \, \bigg[ \widehat{\mathcal{C}}_{AA}^{3, \text{break}} \, \frac{1}{\epsilon^3} + \widehat{\mathcal{B}}_{AA}^{3, \text{break}} \, \frac{1}{\epsilon^2} + \widehat{\mathcal{A}}_{AA}^{3, \text{break}} \, \frac{1}{\epsilon} \bigg] \, \widehat{p}^{2} \, \overline{p}^{\mu}\\
        &+ \frac{e^6}{(16 \pi^2)^3} \, \bigg[ \overline{\mathcal{A}}_{AA}^{3, \text{break}} \, \frac{1}{\epsilon} + \mathcal{F}_{AA}^{3, \text{break}} \bigg] \, \overline{p}^{2} \, \overline{p}^{\mu},
    \end{aligned}
\end{equation}
with $\Delta \equiv \widehat{\Delta}+\Delta_{\text{ct}}$, which determines the full
symmetry-breaking of the gauge boson self-energy at the 3-loop level, 
according
to Eq.\ (\ref{Eq:PerturbativeSTIRequirement}).
Beside the divergent symmetry-breaking contributions, we also obtain 
the finite symmetry-breakings, shown in the last line of 
Eq.\ (\ref{Eq:GaugeBosonSelfEnergy3LoopBreaking})
with coefficient $\mathcal{F}_{AA}^{3, \text{break}}$, as announced
earlier in Sec.\ \ref{Sec:SymmetryRestorationProcedure}.
Note that contracting the result in Eq.\ (\ref{Eq:GaugeBosonSelfEnergy3Loop})
with $p_{\nu}$ identically reproduces the UV-divergent part
of Eq.\ (\ref{Eq:GaugeBosonSelfEnergy3LoopBreaking}); thus, serving 
as a strong consistency check for our results.

Repeating this procedure for all other power-counting
UV-divergent, $\Delta$-operator inserted, 1PI Green functions,
we obtain the full symmetry-breaking $\Delta_{\mathrm{ct}}^{n=3}$ at the
3-loop level, including finite contributions, highlighting
the efficiency of our method.
From this result the finite symmetry-restoring counterterm action 
$S^{n=3}_{\mathrm{fct}}$ can then be obtained via an 
``inverse BRST-transformation'' using Eq.\ (\ref{Eq:DefinitionOfDelta})
and $S^{n}_{\mathrm{ct}}=S^{n}_{\mathrm{sct}}+S^{n}_{\mathrm{fct}}$.

Finally, the complete singular counterterm action for the 
right-handed Abelian chiral gauge theory at the 3-loop level 
is given by
\begin{equation}\label{Eq:Ssct_3-Loop}
    \begin{aligned}
        S^{3}_{\mathrm{sct}} = 
        &\phantom{-\,\,} \frac{e^6}{(16 \pi^2)^3} \,
        \bigg[ \mathcal{B}_{AA}^{3, \text{inv}} \, \frac{1}{\epsilon^2} + \mathcal{A}_{AA}^{3, \text{inv}} \, \frac{1}{\epsilon} \bigg] 
         \, \Dintx \, \Big(-\frac{1}{4} \, \overline{F}^{\mu\nu} \overline{F}_{\mu\nu}\Big)\\
        &- \frac{e^6}{(16 \pi^2)^3} \, 
        \bigg[ \mathcal{C}_{\psi\overline{\psi}, ji}^{3, \text{inv}} \, \frac{1}{\epsilon^3} + \mathcal{B}_{\psi\overline{\psi}, ji}^{3, \text{inv}} \, \frac{1}{\epsilon^2} + \mathcal{A}_{\psi\overline{\psi}, ji}^{3, \text{inv}} \, \frac{1}{\epsilon} \bigg] \\
        &\hspace{1.75cm}\times \Dintx \, \Big( \overline{\psi}_j i \overline{\slashed{\partial}} \mathbb{P}_{\mathrm{R}} \psi_i - e \big(\mathcal{Y}_R\big)_{kj} \overline{\psi}_k \overline{\slashed{A}} \mathbb{P}_{\mathrm{R}} \psi_i \Big)\\
        &- \frac{e^6}{(16 \pi^2)^3} \,
        \bigg[ \widehat{\mathcal{C}}_{AA}^{3, \text{break}} \, \frac{1}{\epsilon^3} + \widehat{\mathcal{B}}_{AA}^{3, \text{break}} \, \frac{1}{\epsilon^2} + \widehat{\mathcal{A}}_{AA}^{3, \text{break}} \, \frac{1}{\epsilon} \bigg] \,
        \Dintx \, \frac{1}{2} \overline{A}_{\mu} \widehat{\partial}^2 \overline{A}^{\mu}\\
        &+ \frac{e^6}{(16 \pi^2)^3} \,
        \overline{\mathcal{A}}_{AA}^{3, \text{break}} \, \frac{1}{\epsilon} \,
        \Dintx \, \frac{1}{2} \overline{A}_{\mu} \overline{\partial}^2 \overline{A}^{\mu}\\
        &- \frac{e^6}{(16 \pi^2)^3} \, 
        \bigg[ \mathcal{B}_{\psi\overline{\psi}, ji}^{3, \text{break}} \, \frac{1}{\epsilon^2} + \mathcal{A}_{\psi\overline{\psi}, ji}^{3, \text{break}} \, \frac{1}{\epsilon} \bigg] \,
        \Dintx \, \Big( \overline{\psi}_j i \overline{\slashed{\partial}}  \mathbb{P}_{\mathrm{R}} \psi_i \Big)\\
        &- \frac{e^8}{(16 \pi^2)^3} \,
        \overline{\mathcal{A}}_{AAAA}^{3, \text{break}} \, \frac{1}{\epsilon} \,
        \Dintx \, \frac{1}{8} \overline{A}_{\mu} \overline{A}^{\mu} \overline{A}_{\nu}  \overline{A}^{\nu}.
    \end{aligned}
\end{equation}
The first three lines represent the usual divergent and symmetric counterterms,
while the subsequent four lines consist of divergent symmetry-restoring counterterms. 
Additionally, the full finite symmetry-restoring counterterm action is provided by
\begin{equation}
    \begin{aligned}
        S^{3}_{\mathrm{fct}} = &\phantom{+\,\,} \frac{e^6}{(16 \pi^2)^3} \,
        \mathcal{F}_{AA}^{3, \text{break}}
        \intx \, \frac{1}{2} \overline{A}_{\mu} \overline{\partial}^2 \overline{A}^{\mu}
        - \frac{e^6}{(16 \pi^2)^3} \, 
        \mathcal{F}_{\psi\overline{\psi}, ji}^{3, \text{break}}
        \intx \, \overline{\psi}_j i \overline{\slashed{\partial}} \mathbb{P}_{\mathrm{R}} \psi_i\\
        &- \frac{e^8}{(16 \pi^2)^3} \,
        \mathcal{F}_{AAAA}^{3, \text{break}}
        \intx \, \frac{1}{8} \overline{A}_{\mu} \overline{A}^{\mu} \overline{A}_{\nu} \overline{A}^{\nu}.
    \end{aligned}
\end{equation}
This finite counterterm action maintains the same structure as at the 
1- and 2-loop level. 
Consequently, unlike the singular counterterm action, no new field 
monomials emerge at the 3-loop level, 
cf.\ results in Refs.\ \cite{Belusca-Maito:2021lnk,Stockinger:2023ndm}.
It is important to note that above, we attributed the breaking of the 
Ward identity associated with the fermion self-energy and the fermion-gauge
boson interaction current entirely to the fermion self-energy. This, however,
is a choice, as it is always possible to add a finite symmetric counterterm
which does neither spoil the Slavnov-Taylor identity, nor change
physics, but just switches between different renormalization schemes.

Currently, we are extending our \texttt{FORM}-based computational setup 
to renormalize this model at the 4-loop level, aiming to advance 
the frontier of multi-loop calculations within the framework of the BMHV scheme.
We expect the structure of the finite symmetry-restoring counterterm action
to remain the same at higher loop-orders, as discussed in Ref.\ \cite{Stockinger:2023ndm}.


\section{Non-Abelian chiral Gauge Theory at the 2-loop Level}\label{Sec:NonAbelianTheories}

In the previous section we have seen the full renormalization of an Abelian toy model at the 3-loop level. Let us now turn to a brief summary of particulars in the non-Abelian case at the multi-loop, i.e.\ 2-loop, level.
The model (with scalars) has already been treated successfully at the 1-loop level in Ref.\ \cite{Belusca-Maito:2021lnk}.

The Lagrangian has the generic form of a Yang-Mills theory with $N_f$ chiral, right-handed fermions and sterile, left-handed partners, comprising the $D$-dimensional fermion, kinetic term,
\begin{equation}\mathcal{L}_\mathrm{kin}^{\mathrm{fermion}}+\mathcal{L}_{\mathrm{int}}^{\mathrm{fermion}}-\frac{1}{4}F^2-\overline{c}\partial D_Ac+\mathcal{L}_{\mathrm{g-fix}}+\mathcal{L}_{\mathrm{ext}}.
\end{equation}
The covariant derivative for fermions and ghosts, respectively, is given by,
\begin{equation}
    D^{\mu}_{ab} = \partial_{\mu}\delta^{a b} +igT_{ab}^{c}G^{c\mu},\qquad  (D_A)^{\mu}_{ab} = \partial_{\mu}\delta^{a b} +gf^{abc}G^{c\mu}.
\end{equation}
As before the simplest prescription for the fermion-gauge boson interaction term is chosen, i.e.,
\begin{equation}
    \mathcal{L}^\mathrm{fermion}_{\mathrm{int}}=-g T^a_{ij}\,\overline{\psi_{Ri}}\overline{\slashed{G}}^a\psi_{Rj}.
\end{equation}

One of the hallmarks of the non-Abelian setting consists in the self-interactions of gauge bosons which, in our setup, lead to interactions of Faddeev-Popov ghosts and BRST transformations $s_D\phi$, for $\phi\in\{\overline{\psi},\psi,G,c\}$, which are non-linear in dynamical fields.
The external fields, which were introduced in Eq.\ (\ref{Eq:SlavnovTaylorIdentity}), couple to these non-linear operators as
\begin{equation}\mathcal{L}_{\text{ext}}=\rho^{\mu}_a s_D G^a_{\mu}+\zeta_a s_D c^a+\overline{R}^{i}s_D\psi_{Ri}+R^is_D \overline{\psi_{Ri}}.\end{equation}
In this way the renormalization of diagrams with BRST operator insertions is automatically included in the renormalization of the Lagrangian.
The BRST operator is fermionic and increases the ghost number by one.
Statistics and quantum numbers of the external fields then follow from the usual requirements of any term in the Lagrangian to be Lorentz invariant, bosonic and of dimension four. 

In general we shall stick to the outline of the symmetry restoration procedure given in Sec.\ \ref{Sec:SymmetryRestorationProcedure}. More concretely we must ensure the validity of Eq.\ (\ref{Eq:PerturbativeSTIRequirement}) which at 2-loop order stipulates a relation between the desired finite 2-loop counterterms, genuine 2-loop $\Delta$-Green functions and Green functions with insertions of 1-loop counterterms or 1-loop symmetry breakings.
There is a list of complications particular to the non-Abelian setting encountered at the 2-loop level, which will be discussed in what follows.


The first complication lies with the structure of appearing symmetry-restoring counterterms $S^1_{\mathrm{fct}}$ and the determination of $\Delta^{1}_{\mathrm{ct}}$. 
As a concrete example, the symmetry breaking requires the existence of 1-loop finite counterterms corresponding to monomials $ S_{\overline{R} c \psi_{R}} $ 
and $ S_{Rc\overline{\psi_R}} $ 
involving the external $R/\overline{R}$-field sources of the fermions. Sample $\Delta$-diagrams and the resulting counterterm Lagrangian can be written as
\begin{center}
\begin{tabular}{c c c c c}
\(\big(\Delta\cdot\Gamma\big)^1\,\,\)
\(\supset \)
\raisebox{-30pt}{\includegraphics[scale=.45]{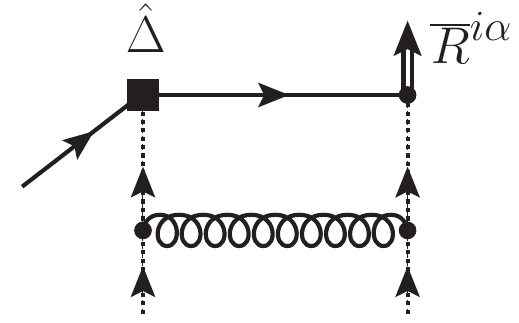}}\(+\)&\(\mathrm{h.c.}\)& \(\Longrightarrow \quad \displaystyle  S_{\text{fct}}^{1}\,\supset\, -g^2 \frac{\xi C_{2}(G)}{4}
\big(S_{\overline{R}c\psi_R}+S_{Rc\overline{\psi}_R}\big)
\)\\
\end{tabular}
\end{center}


To compute the symmetry breaking at the 2-loop level, the operator $\Delta^{\leq1L}$ is required, defined via Eq.\ (\ref{Eq:DefinitionOfDelta}) as
\begin{equation}
    \Delta^{\leq1L}=\widehat{\Delta} + \Delta^{1}_{\mathrm{ct}}=\widehat{\Delta}+b_D(S^{1}_{\text{ct}}).
\end{equation}
The first term corresponds to the familiar tree-level breaking which we encountered in the Abelian model (cf.\ Eq.\ (\ref{DeltahatDef})) but for the diagonal hypercharge matrix replaced by a $SU(N)$ generator.
The second term contains the operator $b_D$, which is the non-Abelian generalization of $s_D$,
\begin{equation}\label{bDDef}
b_D=s_D+\int d^Dx\,\frac{\delta S_0}{\delta G^{a\mu}}\frac{\delta}{\delta\rho^{a\mu}}+\frac{\delta S_0}{\delta \psi_{j\beta}}\frac{\delta}{\delta \overline{R}_{j\beta}}+\frac{\delta S_0}{\delta \overline{\psi}_{i\alpha}}\frac{\delta}{\delta R_{i\alpha}}+\frac{\delta S_0}{\delta c^a}\frac{\delta}{\delta\zeta^a}.
\end{equation}
It is clear that in the Abelian case the operator simplifies to the BRST operator $s_D$ since loop corrections of the counterterm action are manifestly independent of external field pieces, but in the non-Abelian case the external field terms contribute.

In the evaluation of $\Delta^1_{\mathrm{ct}}$ from the counterterms involving the $R$-field, the $b_D$ operator produces additional evanescent correction terms which are important for consistency at higher loops. This is so because the fermion kinetic term, which enters via the tree level action $S_0$ in Eq.\ (\ref{bDDef}), is fully $D$-dimensional,
\begin{equation}
    b_{D}\big(S_{\overline{R} c \psi_{R}}\big)=\int d^{D}x \,\frac{\delta S_0}{\delta \psi_{i\alpha}}\frac{\delta S_{\overline{R} c \psi_{R}}}{\delta\overline{R}_{i\alpha}}+\ldots=\overline{\Delta}^{1L, c\overline{\psi}\psi}_{\mathrm{fct}}+\widehat{\Delta}^{1L, c\overline{\psi}\psi}_{\mathrm{fct}}+\dots
\end{equation}
The appearance of such an evanescent term entering $\Delta_{\mathrm{ct}}^1$ is somewhat counterintuitive and cannot happen in the Abelian case.

The second big complication has to do with the practical implementation not only of the $\Delta$-operator, but also the external field pieces. 
As heralded in Sec.\ \ref{Sec:ComputationalSetup}, we have built an automated setup in \texttt{FeynArts} which we briefly sketch here.
The $\Delta$-vertex is fermionic but is effectively treated as a Lagrangian term, which should be bosonic. We couple the operator to an auxiliary anti-ghost $\overline{\omega}_{\Delta}^a$ such that
\begin{equation}
    \Delta\cdot\Gamma_{\phi_1\dots\phi_n}=\frac{\delta\Gamma_{\phi_1\dots\phi_n}}{\delta \overline{\omega}_{\Delta}^a}\bigg\vert_{\overline{\omega}_{\Delta}^a=0}.
\end{equation}
BRST sources such as $\rho^{a\mu}$ or $R^{j\beta}$
have fermionic and bosonic statistics, while transforming as a Lorentz vector and spinor, respectively. In the \texttt{FeynArts} implementation they are modelled as composite fields comprised of auxiliary anti-ghost fields $\overline{\omega}$ and appropriate quantum fields as
\begin{equation}
\rho^{a\mu}\longrightarrow\overline{\omega}_{\rho}^aA^{\mu}\qquad R_{j\beta}\longrightarrow\overline{\omega}_{R}\chi_{j\beta}.
\end{equation}
Since in \texttt{FeynArts} vertices of four fermionic objects are difficult to handle, it is advantageous to split the vertices using auxiliary scalar fields as in the following sketch of the vertex $\widehat{\Delta}$ which should contain four fermions $\overline{\omega}_{\Delta}^a$, $c^a$, $\psi$, $\overline{\psi}$,

\begin{center}
\begin{tabular}{c c c c c}
\includegraphics[scale=.4]{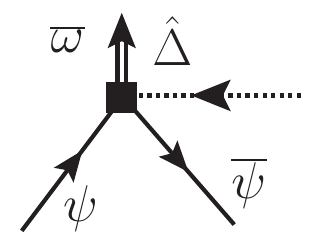}&
\raisebox{22pt}{$\Longrightarrow$}
&
\includegraphics[scale=.4]{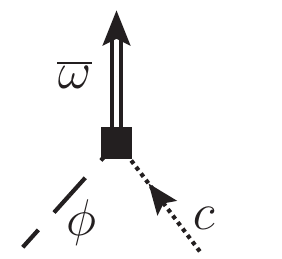}&
\raisebox{22pt}{\(\!\!\!\!\otimes\)} &
\includegraphics[scale=.4]{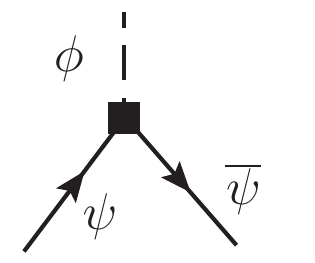}
\end{tabular}
\end{center}
In this way all diagrams required for the renormalization procedure, including diagrams with external sources and/or $\Delta$-vertices, can be implemented in \texttt{FeynArts}.

An additional complication compared to the Abelian case is the existence of new Green functions, some of which are non-vanishing for the first time at the 2-loop level, e.g.\
trilinear ghost Green functions, quintic breaking of the gluonic sector and finite contributions of all of the BRST sources. Sample diagrams are

\begin{center}
\begin{tabular}{c c c}
\includegraphics[scale=.4]{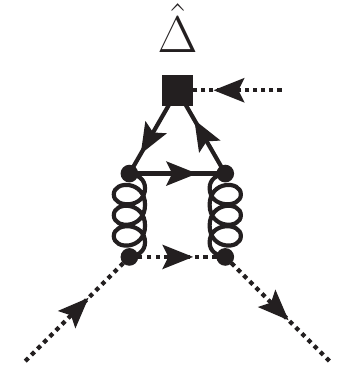}
&
\qquad\includegraphics[scale=.4]{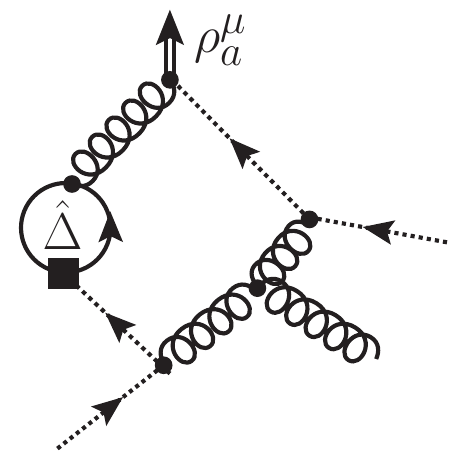}
&
\qquad\includegraphics[scale=.4]{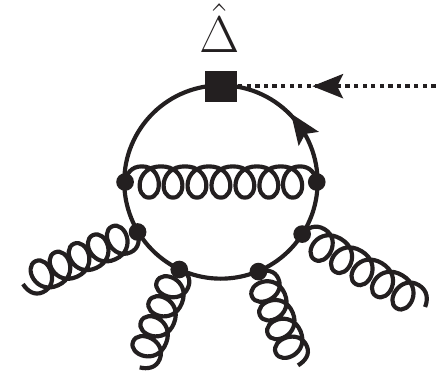}
\end{tabular}
\end{center}

As a final complication, the general structure of Slavnov-Taylor identities between Green functions is significantly more complicated than in the Abelian case and all ingredients of the LHS and the RHS of the quantum action principle (Eq.\ (\ref{Eq:QAPofDReg})) become more involved.
In the following we exemplify the complexity of the required computations but also describe a multitude of checks which have been successfully performed.

By virtue of Eq.\ (\ref{Eq:QAPofDReg}), the symmetry breaking can be determined in two ways as outlined in Sec.\ \ref{Sec:SymmetryRestorationProcedure}.
Therefore comparing both approaches provides a means of checking the implementations for consistency. On the one hand, the LHS of Eq.\ (\ref{Eq:QAPofDReg}) features products of standard Green functions. On the other hand, the RHS consists of 2-loop $\widehat{\Delta}$-insertions and $\Delta^{1}_{\mathrm{ct}}$-corrections. Further, for the examples discussed below, the loop calculation can be done either 
straightforwardly, including complete finite contributions, 
via \texttt{TARCER} \cite{Mertig:1998vk} or using the tadpole decomposition method discussed in Sec.\ \ref{Sec:ComputationalSetup}. In total there are thus four different ways to compute the symmetry breaking via the LHS or the RHS of Eq.\ (\ref{Eq:QAPofDReg}). We find full agreement.

In the following we briefly discuss two concrete examples of the symmetry breaking and symmetry restoration in the non-Abelian case at the 2-loop level where all the mentioned checks have been done. 
The simplest and most straightforward case is the Slavnov-Taylor identity corresponding to the transversality of the gluon self-energy. The corresponding version of Eq.\ (\ref{Eq:QAPofDReg}) reads 
\begin{equation}
    \Gamma^{\text{DRen}}_{\rho^{\mu}_a c_b}\cdot\Gamma^{\text{DRen}}_{G^{\mu}_a G^{\nu}_b}=\Delta\cdot\Gamma^{\text{DRen}}_{c_a G^{\nu}_b}
\end{equation}
and the LHS contains $p^{\mu}\Gamma_{G^{b\nu}G^{a\mu}}^{\mathrm{DRen},\,\mathrm{fin}}$ and thus describes the violation of transversality.
An excerpt of diagrams includes,\vspace{.5cm}

\begin{tabular}{c c c c c}
    \raisebox{-12pt}{\(\displaystyle\includegraphics[scale=.4]{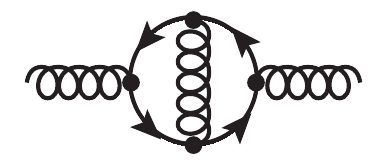}\)}&+&
    \raisebox{-32pt}{\(\displaystyle\includegraphics[scale=.4]{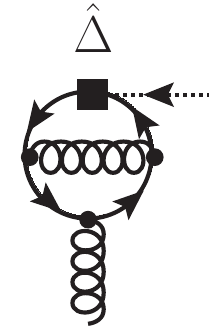}\)}&+&
    \(\displaystyle \dots\;\text{\textit{Abelian Diagrams}}\)\\
    \raisebox{-12pt}{\(\displaystyle\includegraphics[scale=.4]{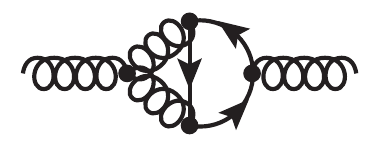}\)}&+&
    \raisebox{-32pt}{\(\displaystyle\includegraphics[scale=.4]{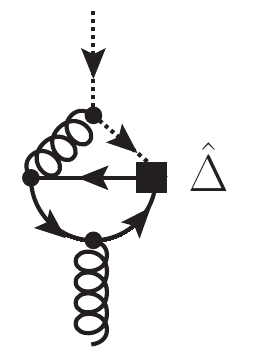}\)}&+&
    \raisebox{-5pt}{\(\displaystyle\includegraphics[scale=.4]{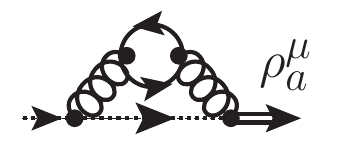}\)}\\
    \text{\textit{Non-Abelian Diagrams}}&\hphantom{+}&
    \textit{Internal}$\;\widehat{\Delta}$-\textit{\large{vertex}}&\hphantom{+}&
    \text{\textit{external source}}\\\textit{\large{ Green function}}\\

\end{tabular}
\vspace{.5cm}

Interestingly, owing to the non-Abelian interactions, the $\Delta$-vertex can be internal to the diagram contrary to the Abelian case. These diagrams can be computed to obtain the symmetry breaking including the cross-checks mentioned above, and ultimately to obtain the symmetry-restoring counterterms. In this case, transversality can be restored by the following finite counterterm action,
\begin{equation}
S^{(2)}_{\mathrm{fct}}\supset\frac{g^4}{256\pi^4}\Big(-\frac{7}{216} C_A + \frac{11}{96} C_F\Big)\overline{G}^{a}_{\mu}\overline{\square}\,\overline{G}^{a\mu}.
\end{equation}
As a further cross-check, the term proportional to $C_F$ corresponds to Abelian contributions and agrees with the 2-loop result in the Abelian model, see Refs.\ \cite{Belusca-Maito:2021lnk,Belusca-Maito:2022usw,Stockinger:2023ndm}.

For a second, more involved example we shall turn to the non-Abelian counterpart of the famous Abelian Ward identity of the electron self-energy and vertex-correction. This relation is significantly complicated by the appearance of prefactors containing Green functions with $R/\overline{{R}}$- and $\rho^\mu$-sources. The LHS and RHS of the corresponding version of Eq.\ (\ref{Eq:QAPofDReg}), including sample diagrams, reads \vspace{.5cm}

    \begin{tabular}{c c c}
\(\displaystyle \Gamma^{\text{DReg}}_{c_a\rho^{\mu}_b}\cdot\Gamma^{\text{DReg}}_{\psi_{j\beta}\overline{\psi}_{i\alpha}G^{\mu}_b}\) & + & \(\displaystyle\Gamma^{\text{DReg}}_{\psi_{j\beta}c_a\overline{R}_{l\delta}}\cdot\Gamma^{\text{DReg}}_{\psi_{l\delta}\overline{\psi}_{i\alpha}}\)\\
\(\displaystyle \delta^{ab}p^{\mu}\begin{gathered}\includegraphics[scale=.4]{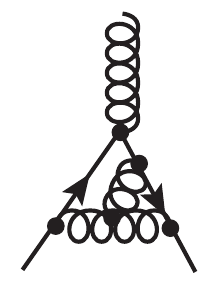}\end{gathered}-igT^b_{ij}\overline{\gamma}^{\mu}\mathbb{P}_R\begin{gathered}\includegraphics[scale=.4]{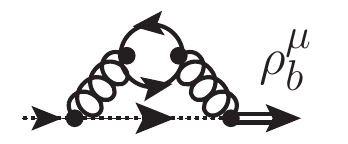}\end{gathered}\dots\)& \(\displaystyle+\) &
\(\displaystyle\dots\begin{gathered}\includegraphics[scale=.4]{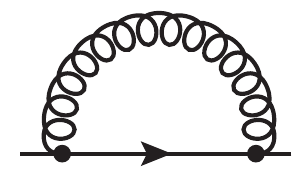}\end{gathered}\quad\otimes\begin{gathered}\includegraphics[scale=.4]{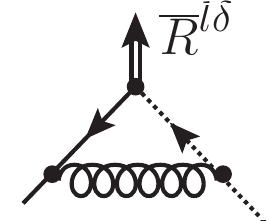}\end{gathered}+\dots\)\\
\(\displaystyle\Gamma^{\text{DReg}}_{\psi_{j\beta}c_a\overline{R}_{l\delta}}\cdot\Gamma^{\text{DReg}}_{\psi_{l\delta}\overline{\psi}_{i\alpha}}\) & \(\displaystyle =\) & \(\displaystyle
\Delta\cdot\Gamma^{\text{DReg}}_{\psi_{j\beta}\overline{\psi}_{i\alpha}c_a}\)\\
similar contribution&\hphantom{=}&
\(\displaystyle\dots+
\begin{gathered}\includegraphics[scale=.4]{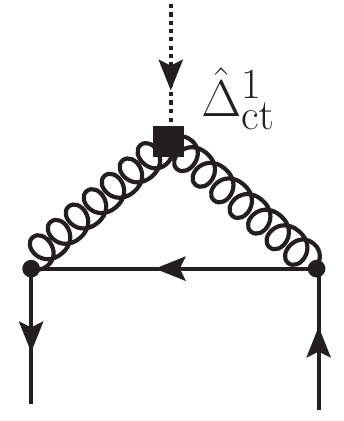}\end{gathered}+\dots\)\\
\end{tabular}

\vspace{.5cm}

Again, all described cross-checks have been carried out.
We may define a finite local counterterm action which upon action of $b_D$ cancels this symmetry breaking,
\begin{equation}
    S^{(2)}_{\mathrm{fct}}\supset\frac{g^4}{256\pi^4}\Big(-\frac{4}{3}C_A^2+\frac{55}{432}C_A+\frac{1121}{216}C_AC_F-\frac{127}{36}C_F^2-\frac{41}{216}C_F\Big)(i\overline{\psi_{\mathrm{R}}}_i\overline{\slashed{\partial}}\psi_{\mathrm{R}_i}).
\end{equation}
Again the Abelian limit is consistent with our previous results.
We note in passing that since the action of the $b_D$ operator on (finite) counterterms mixes different contributions of breaking Green functions, this action piece would similarly affect e.g.\ $(\Delta\cdot\Gamma)_{\psi\overline{R}cc}$ and   $(\Delta\cdot\Gamma)_{G\psi\overline{\psi}c}$.
The ultimate determination of the finite counterterm action is thus complicated by these consistency conditions.


\section{Exploring BMHV in the Standard Model Context: Conclusions and Perspectives}\label{Sec:ExplorationsAndSM}

In the previous sections we have applied the BMHV scheme to chiral gauge theories with simple or Abelian gauge group. We have shown that the scheme can be applied, the symmetry breaking can be determined, and the required symmetry-restoring counterterms can be found and have a rather simple structure. In the following we discuss several issues which affect the application of the BMHV scheme to the SM. 

Apart from the Higgs vacuum expectation value and spontaneous electroweak symmetry breaking, major intricacies result from the mixing of left- and right-handed fermions. To illustrate the issues, we focus on electron fields $\psi_{L,R}$ which couple to a hypercharge gauge boson $B^\mu$ with different $U(1)$ hypercharges ${\cal Y}_{L,R}$ but which have the same electric charge. The question is how to regularize the kinetic and gauge interaction terms. We discuss three options. 

The first, straightforward option is to define a full electron Dirac spinor  $\psi=\psi_L+\psi_R$ 
and write the regularized Lagrangian for kinetic and gauge interactions as (see Ref.\ \cite{Cornella:2022hkc})
\begin{equation}\label{Eq:SMfermion1}
    \begin{aligned}
        \underbrace{
        \overline{\psi} i \slashed{\partial} \psi
        }_{=
        \overline{\psi_L}i\overline{\slashed{\partial}}\psi_L+
        \overline{\psi_R}i\widehat{\slashed{\partial}}\psi_L+\ldots}
        \hspace{-1cm}
        - g {\cal Y}_L
        \overline{\psi_L} \overline{\slashed{B}}\psi_L
        - g {\cal Y}_R
        \overline{\psi_R} \overline{\slashed{B}}\psi_R
    \end{aligned}
\end{equation}
Here, like in Eq.\ (\ref{Eq:LfermionsR}), the kinetic term must be $D$-dimensional, but the interaction term has been chosen to be purely 4-dimensional. 
As always in the BMHV scheme, local gauge invariance and BRST invariance are broken. However, Eq.\ (\ref{Eq:SMfermion1}) also breaks global hypercharge invariance: the $\widehat{\slashed{\partial}}$-part involves a product between $\overline{\psi_R}$ and $\psi_L$ whose hypercharges do not add up to zero. This is different in Eq.\ (\ref{Eq:LfermionsR}), where $\psi_L$ is a fictitious sterile field which can be assumed to have the same global hypercharge as $\psi_R$. 

A second option is to modify the gauge interaction term to become also $D$-dimensional,
\begin{align}
    \overline{\psi} i\slashed{\partial}\psi
  - g {\cal Y}_L
  \overline{\psi_L} \overline{\slashed{B}}\psi_L
  - g {\cal Y}_R
  \overline{\psi_R} \overline{\slashed{B}}\psi_R
  - g {\cal Y}_{RL}
  \overline{\psi_R} \widehat{\slashed{B}}\psi_L
  - g {\cal Y}^{\dagger}_{RL}
  \overline{\psi_L} \widehat{\slashed{B}}\psi_R
\end{align}
A motivation for this additional interaction term comes from comparing to the standard $D$-dimensional treatment of QED. There, both the kinetic and the interaction term is treated fully $D$-dimensionally and gauge invariance is manifestly preserved. One may hope that the additional interaction term (for suitable choice of the coefficient ${\cal Y}_{RL}$) improves the symmetry properties of the scheme, particularly with respect to the photon interactions which are part of the SM.

A third option is to repeat the treatment of the simpler toy models and introduce fictitious sterile fermion partners for all physical fermions. 
For the electron we would then introduce two Dirac fermions
  $\psi_1=\psi_L+\psi^{\text{st}}_R$, 
  $\psi_2=\psi_R+\psi^{\text{st}}_L$ where $\psi^{\text{st}}_{L,R}$ are the fictitious partners.  Then the $D$-dimensional Lagrangian can be written as
\begin{align}
\mathcal{L}_\text{fermions} = 
\left[\overline{\psi_1} i\slashed{\partial}\psi_1
  - g {\cal Y}_L
  \overline{\psi_L} \overline{\slashed{B}}\psi_L\right]
  + \left[ \overline{\psi_2} i\slashed{\partial}\psi_2
  - g {\cal Y}_R
  \overline{\psi_R} \overline{\slashed{B}}\psi_R\right]
\end{align}
Here each square bracket is completely analogous to Eq.\ (\ref{Eq:LfermionsR}), and the hypercharges of the sterile partners can be chosen such that global hypercharge invariance is not broken. Such a treatment has been used in Ref.\ \cite{Martin:1999cc}. Apart from appearing non-economical it leads to a complicated mixing propagator system once a mass term $m\overline{\psi_R}\psi_L$ is added.

In summary, each of the options has advantages and disadvantages, and explicit calculations will determine which option is most efficient in practice and/or leads to the simplest results. Such explicit investigations are ongoing.
To conclude, based on recent developments on different frontiers of $\gamma_5$,
including the number of loops crucial for high-precision physics, 
the treatment of non-Abelian chiral gauge theories
essential for both SM and BSM applications,
as well as other intricate aspects of the Standard Model,
such as those discussed above, 
we are confident to provide a comprehensive and rigorous SM renormalization 
in the BMHV scheme.
This will encompass the incorporation of symmetry-restoring counterterms, 
which will be made available to practitioners in the near future.


\section*{Acknowledgments}

P.K., D.S.\ and M.W.\ acknowledge financial support by the German Science 
Foundation DFG, grant STO 876/8-1. 
We would like to thank our collaborators 
Herm\`es B\'elusca-Ma\"\i{}to, Paul Ebert, Amon Ilakovac, and Marija Ma\dj{}or-Bo\v{z}inovi\'c,
as well as Andreas von Manteuffel for insightful ideas 
and valuable discussions. 

\appendix

\section{Abelian chiral Gauge Theory: Explicit 3-loop Coefficients}\label{App:3-Loop-Coeffs}

Here, we present explicit results for the 3-loop coefficients
introduced in Sec.\ \ref{Sec:AbelianMultiLoop}.
We start with the coefficients for the purely gauge bosonic terms:

\vspace{0.3cm}\noindent\textbf{Gauge Boson 3-Loop Coefficients:}
\begin{align}
\begin{split}
    \mathcal{B}_{AA}^{3, \text{inv}} &= \frac{4}{162} \Big( 3 \, \text{Tr}\big(\mathcal{Y}_R^6\big) - 5 \, \text{Tr}\big(\mathcal{Y}_R^4\big) \text{Tr}\big(\mathcal{Y}_R^2\big) \Big)\label{Eq:BphotonInv}
\end{split}\\[1.5ex]
\begin{split}
    \mathcal{A}_{AA}^{3, \text{inv}} &= - \frac{1}{1620} \Big( 2552 \, \text{Tr}\big(\mathcal{Y}_R^6\big) + 61 \, \text{Tr}\big(\mathcal{Y}_R^4\big) \text{Tr}\big(\mathcal{Y}_R^2\big) \Big)\label{Eq:AphotonInv}
\end{split}\\[1.5ex]
\begin{split}
    \widehat{\mathcal{C}}_{AA}^{3, \text{break}} &= \frac{1}{18} \, \text{Tr}\big(\mathcal{Y}_R^6\big)\label{Eq:CphotonBreakEvan}
\end{split}\\[1.5ex]
\begin{split}
    \widehat{\mathcal{B}}_{AA}^{3, \text{break}} &= - \frac{1}{1080} \Big( 529 \, \text{Tr}\big(\mathcal{Y}_R^6\big) + 122 \, \text{Tr}\big(\mathcal{Y}_R^4\big) \text{Tr}\big(\mathcal{Y}_R^2 \Big)\label{Eq:BphotonBreakEvan}
\end{split}\\[1.5ex]
\begin{split}
    \widehat{\mathcal{A}}_{AA}^{3, \text{break}} &= \frac{1}{64800} \Big( \big( 156672 \, \zeta_{3} - 49427 \big) \text{Tr}\big(\mathcal{Y}_R^6\big) - 8374 \, \text{Tr}\big(\mathcal{Y}_R^4\big) \text{Tr}\big(\mathcal{Y}_R^2\big) \Big)\label{Eq:AphotonBreakEvan}
\end{split}\\[1.5ex]
\begin{split}
    \overline{\mathcal{A}}_{AA}^{3, \text{break}} &= \frac{1}{1080} \Big( 18 \, \text{Tr}\big(\mathcal{Y}_R^6\big) + 79 \, \text{Tr}\big(\mathcal{Y}_R^4\big) \text{Tr}\big(\mathcal{Y}_R^2\big) \Big)\label{Eq:AphotonBreakNonEvan}
\end{split}\\[1.5ex]
\begin{split}\label{Eq:FPhotonBreak}
    \mathcal{F}_{AA}^{3, \text{break}} &= - \frac{1}{21600} \Big( \big( 35242 + 8448 \, \zeta_{3} \big) \text{Tr}\big(\mathcal{Y}_R^6\big) + 1639 \, \text{Tr}\big(\mathcal{Y}_R^4\big) \text{Tr}\big(\mathcal{Y}_R^2\big) \Big)
\end{split}\\[1.5ex]
\begin{split}\label{Eq:AQuarticPhotonBreak}
    \overline{\mathcal{A}}_{AAAA}^{3, \text{break}} &= \frac{1}{54} \Big( 6 \, \text{Tr}\big(\mathcal{Y}_R^8\big) + 13 \, \text{Tr}\big(\mathcal{Y}_R^6\big) \text{Tr}\big(\mathcal{Y}_R^2\big) + 48 \, \big(\text{Tr}\big(\mathcal{Y}_R^4\big)\big)^2 \Big)
\end{split}\\[1.5ex]
\begin{split}\label{Eq:FQuarticPhotonBreak}
    \mathcal{F}_{AAAA}^{3, \text{break}} &= -\frac{1}{54} \bigg( \frac{1387+2592\,\zeta_{3}}{10} \, \text{Tr}\big(\mathcal{Y}_R^8\big) +  \frac{101}{20} \, \text{Tr}\big(\mathcal{Y}_R^6\big) \text{Tr}\big(\mathcal{Y}_R^2\big) + 51 \, \big(\text{Tr}\big(\mathcal{Y}_R^4\big)\big)^2 \bigg)
\end{split}
\end{align}
Continuing with the coefficients for terms that contain fermions:

\vspace{0.3cm}\noindent\textbf{Fermion 3-Loop Coefficients:}
\begin{align}
\begin{split}\label{Eq:CfermionInv}
    \mathcal{C}_{\overline{\psi}\psi, \, ij}^{3, \text{inv}} &= \frac{1}{6} \big(\mathcal{Y}_R^6\big)_{ij}
\end{split}\\[1.5ex]
\begin{split}\label{Eq:BfermionInv}
    \mathcal{B}_{\overline{\psi}\psi, \, ij}^{3, \text{inv}} &= \frac{1}{324} \Big( 432 \big(\mathcal{Y}_R^6\big)_{ij} - 186 \big(\mathcal{Y}_R^4\big)_{ij} \text{Tr}\big(\mathcal{Y}_R^2\big)\\
    &\hspace{1.3cm} - 6 \big(\mathcal{Y}_R^2\big)_{ij} \text{Tr}\big(\mathcal{Y}_R^4\big) - \big(\mathcal{Y}_R^2\big)_{ij} \big(\text{Tr}\big(\mathcal{Y}_R^2\big)\big)^2 \Big)
\end{split}\\[1.5ex]
\begin{split}\label{Eq:AfermionInv}
    \mathcal{A}_{\overline{\psi}\psi, \, ij}^{3, \text{inv}} &= \frac{1}{3888} \bigg[ 21843 \big(\mathcal{Y}_R^6\big)_{ij} - 4338 \big(\mathcal{Y}_R^4\big)_{ij} \text{Tr}\big(\mathcal{Y}_R^2\big)\\
    &\hspace{1.3cm} - \Big( 2166 \text{Tr}\big(\mathcal{Y}_R^4\big) - 91 \big(\text{Tr}\big(\mathcal{Y}_R^2\big)\big)^2 \Big) \big(\mathcal{Y}_R^2\big)_{ij} + 2430 \text{Tr}\big(\mathcal{Y}_R^5\big) \big(\mathcal{Y}_R\big)_{ij} \bigg]
\end{split}\\[1.5ex]
\begin{split}\label{Eq:BfermionBreak}
    \mathcal{B}_{\overline{\psi}\psi, \, ij}^{3, \text{break}} &= - \frac{1}{3} \bigg[ \big(\mathcal{Y}_R^6\big)_{ij} - \frac{1}{2} \big(\mathcal{Y}_R^4\big)_{ij} \text{Tr}\big(\mathcal{Y}_R^2\big) + \frac{\big(\mathcal{Y}_R^2\big)_{ij}}{54} \Big( 3 \text{Tr}\big(\mathcal{Y}_R^4\big) + 13 \big(\text{Tr}\big(\mathcal{Y}_R^2\big)\big)^2 \Big) \bigg]
\end{split}\\[1.5ex]
\begin{split}\label{Eq:AfermionBreak}
    \mathcal{A}_{\overline{\psi}\psi, \, ij}^{3, \text{break}} &= - \frac{1}{18} \bigg[ 79 \big(\mathcal{Y}_R^6\big)_{ij} - \frac{169}{6} \, \big(\mathcal{Y}_R^4\big)_{ij} \text{Tr}\big(\mathcal{Y}_R^2\big)\\
    &\hspace{1.3cm} - \frac{\big(\mathcal{Y}_R^2\big)_{ij}}{108} \Big( 159 \text{Tr}\big(\mathcal{Y}_R^4\big) - 113 \big(\text{Tr}\big(\mathcal{Y}_R^2\big)\big)^2 \Big) + \frac{45}{4} \, \big(\mathcal{Y}_R\big)_{ij} \text{Tr}\big(\mathcal{Y}_R^5\big) \bigg]
\end{split}\\[1.5ex]
\begin{split}\label{Eq:FfermionBreak}
    \mathcal{F}_{\overline{\psi}\psi, \, ij}^{3, \text{break}} &= - \bigg( \frac{775}{54} + \frac{58}{9} \, \zeta_{3} \bigg) \big(\mathcal{Y}_R^6\big)_{ij} + \frac{10}{9} \, \big(\mathcal{Y}_R^4\big)_{ij} \text{Tr}\big(\mathcal{Y}_R^2\big)\\
    &\quad - \big(\mathcal{Y}_R^2\big)_{ij} \Bigg[ \bigg( \frac{9725}{3888} + \frac{14}{3} \, \zeta_{3} \bigg) \text{Tr}\big(\mathcal{Y}_R^4\big) - \frac{1993}{23328} \, \big(\text{Tr}\big(\mathcal{Y}_R^2\big)\big)^2 \Bigg]\\
    &\quad + \big(\mathcal{Y}_R\big)_{ij} \bigg( \frac{215}{96} - 7 \, \zeta_3 \bigg) \text{Tr}\big(\mathcal{Y}_R^5\big)
\end{split}
\end{align}


\bibliographystyle{unsrt}
\bibliography{bibliography}

% ...


\end{document}